\documentclass{ptephy_v1}

\preprintnumber{XXXX-XXXX} 

\usepackage{xcolor}
\usepackage{siunitx}
\usepackage{amsmath}
\usepackage{mathtools}
\usepackage{empheq}
\usepackage{here}
\usepackage{url}

\begin{document}

\title{Pointing calibration of GroundBIRD telescope using Moon observation data}


\author[1]{Y. Sueno}

\author[2,3]{J.J.A. Baselmans}

\author[2]{A.H.M. Coppens}

\author[4,5]{R.T G\'{e}nova-Santos}

\author[6]{M. Hattori}

\author[7,8]{S. Honda}

\author[2]{K. Karatsu}

\author[9]{H. Kutsuma}

\author[10]{K. Lee}

\author[11]{T. Nagasaki}

\author[12]{S. Oguri}

\author[13]{C. Otani}

\author[4,5,14]{M. Peel}

\author[1]{J. Suzuki}

\author[1]{O. Tajima}

\author[6]{T. Tanaka}

\author[6]{M. Tsujii}

\author[2]{D.J. Thoen}

\author[10]{E. Won}

\affil[1]{Department of Physics, Graduate School of Science, Kyoto University, Kitashirakawa-Oiwakecho, Sakyo-ku, Kyoto 606-8502, Japan \email{sueno.yoshinori.83x@st.kyoto-u.ac.jp}}

\affil[2]{SRON – Netherlands Institute for Space Research, Niels Bohrweg 4, 2333 CA Leiden, The Netherlands}

\affil[3]{Department of Microelectronics, Faculty of Electrical Engineering, Mathematics and Computer Science (EEMCS), Delft University of Technology, Mekelweg 4, 2628 CD Delft, The Netherlands}

\affil[4]{Instituto de Astrof\'{i}sica de Canarias, E-38205 La Laguna, Tenerife, Spain}

\affil[5]{Departamento de Astrofísica, Universidad de La Laguna, Santa  Cruz de Tenerife, E-38206 La Laguna, Spain}

\affil[6]{Astronomical Institute, Graduate School of Science, Tohoku University, 6-3, Aramaki Aza-Aoba, Aoba-ku, Sendai, 980-8578, Japan}

\affil[7]{Division of Physics, Faculty of Pure and Applied Sciences, University of Tsukuba, Ibaraki, 305-8571, Japan}

\affil[8]{Tomonaga Center for the History of the Universe (TCHoU), University of Tsukuba,  Ibaraki, 305-8571, Japan}

\affil[9]{Department of Applied Physics, Graduate School of Engineering, Tohoku University, 6-6-05, Aramakiaoba, Aoba-ku, Sendai-shi, Miyagi, 980-8579, Japan}

\affil[10]{Department of Physics, Korea University, 145 Anam-ro, Seongbuk-gu, Seoul, 02841, Republic of Korea}

\affil[11]{The High Energy Accelerator Research Organization (KEK), 1-1 Oho, Tsukuba, Ibaraki 305-0801, Japan}

\affil[12]{Institute of Space and Astronautical Science (ISAS), Japan Aerospace Exploration Agency (JAXA)
3-1-1 Yoshinodai, Chuo-ku, Sagamihara, Kanagawa 252-5210, JAPAN}

\affil[13]{Terahertz Sensing and Imaging Research Team, RIKEN 519-1399 Aramaki-Aoba, Aoba-ku, Sendai, 980-0845, Miyagi, Japan}

\affil[14]{Imperial College London, Blackett Lab, Prince Consort Road, London SW7 2AZ, UK}

\begin{abstract}

Understanding telescope pointing (i.e., line of sight) is important for observing the cosmic microwave background (CMB) and astronomical objects.
The Moon is a candidate astronomical source for pointing calibration.
Although the visible size of the Moon ($\ang{;30}$) is larger than that of the planets, we can frequently observe the Moon once a month with a high signal-to-noise ratio.
We developed a method for performing pointing calibration using observational data from the Moon.
We considered the tilts of the telescope axes as well as the encoder and collimation offsets for pointing calibration.
In addition, we evaluated the effects of the nonuniformity of the brightness temperature of the Moon, which is a dominant systematic error.
As a result, we successfully achieved a pointing accuracy of $\ang{;3.3}$.
This is one order of magnitude smaller than an angular resolution of $\ang{;36}$.
This level of accuracy competes with past achievements in other ground-based CMB experiments using observational data from the planets.

\end{abstract}

\subjectindex{xxxx, xxx}

\maketitle

\section{Introduction}

Precise measurements of the cosmic microwave background (CMB) provide us with knowledge about the beginning of the universe, such as cosmic inflation prior to the Big Bang~\cite{Kami, Zal}.
An important predictor of the cosmic inflation is the existence of primordial gravitational waves~\cite{Staro}.
Primordial gravitational waves imprint weak odd-parity patterns ($B$-modes) in  the CMB polarization~\cite{Brout, Starobinsky, Kazanas, Sato, Guth, Linde, Albrecht}.
Recent CMB experiments have focused on detecting $B$-modes from primordial gravitational waves, which are expected to be very faint.
Therefore, it is important to establish calibration methods and obtain sufficient statistics by implementing a large number of detectors.
In particular, the calibration of the line-of-sight of each detector (hereafter referred to as pointing) is essential.
This is because time-ordered data (raw data) cannot be converted into map data (patterns in the sky coordinates) without pointing information.
In previous studies, the primary calibration sources used were planets ~\cite{PB_point, QUIET_point, Planck_point}.
Given their small size compared with the large beam widths of CMB telescopes, they can be safely considered point sources.
This is because we do not need to consider possible non-uniformities of them.
For ground-based CMB experiments, the previous study achieved a ratio of pointing accuracy to their beam width is 0.13--0.14~\cite{PB,QUIET_point}.
This is because the pointing error is related to the beam width and leads to a smearing of the beam.
The beam is defined as the telescope response in relation to  the angle from the pointing direction. The angular resolution is represented by the beam width, which is the full width at half maximum (FWHM).

\begin{figure}[t!]
  \centering
  \includegraphics[width = 15cm]{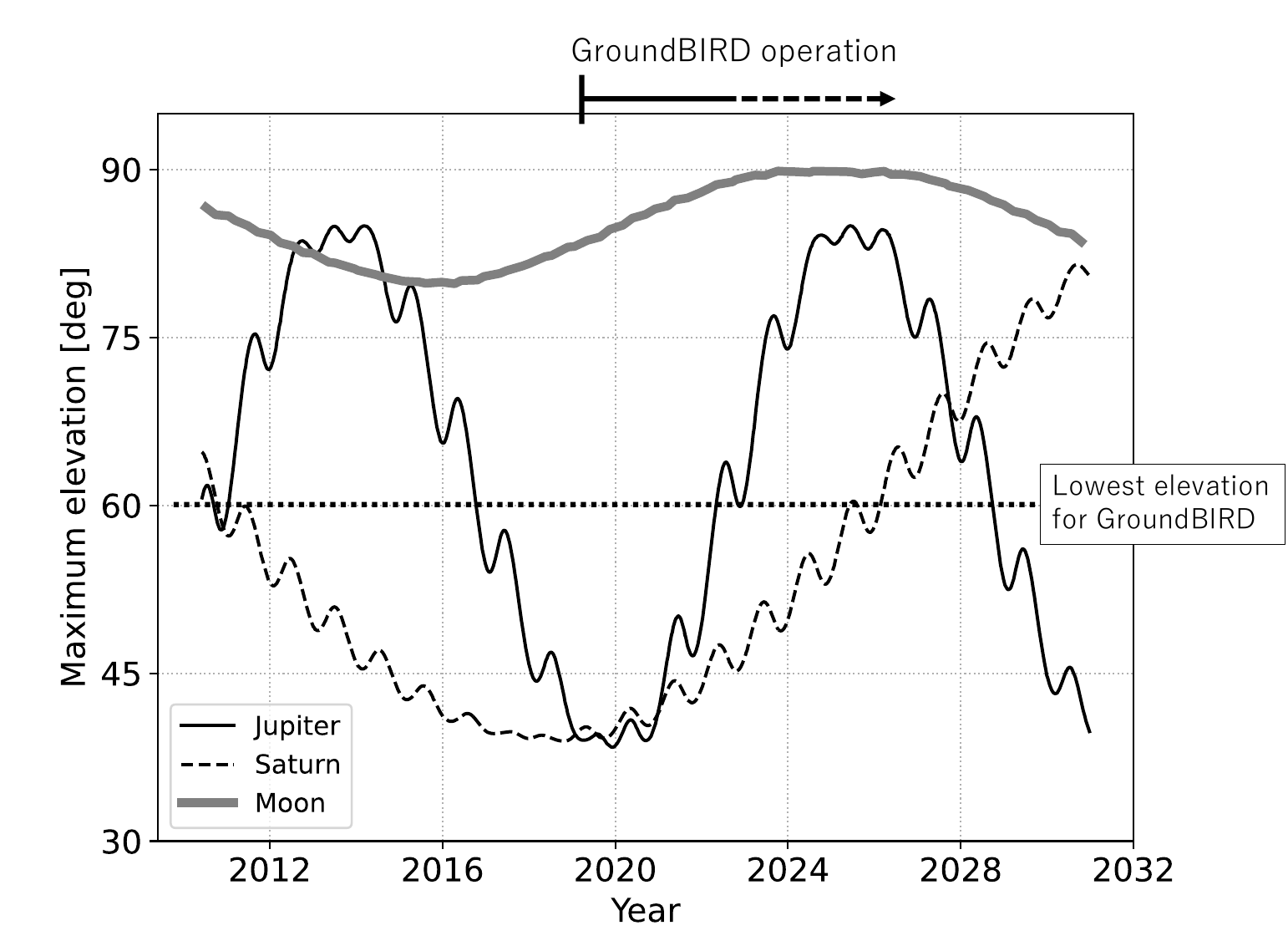}
  \caption{Maximum elevations of Jupiter, Saturn, and the Moon between 2011 and 2031 at the Teide Observatory in the Canary Islands, Spain. 
  The orbital periods of Jupiter and Saturn are 12 years and over 20 years, respectively.
  In the years around 2020, their maximum elevations were lower than $60^{\circ}$.
  However, the Moon has always had a high elevation throughout these years.}
  \label{jupiter}
\end{figure}

Jupiter and Saturn are popular pointing calibration sources for observations in the millimeter wavelength range.
Their positions in the sky were calculated precisely using \textsf{astropy}~\cite{astropy2013, astropy2018}.
Their visible sizes ($\ang{;0.2} \sim \ang{;0.3}$) are sufficiently small compared with the beam width, which is typically $\ang{;1}$ to $\ang{1}$ for CMB telescopes.
The maximum elevation of the planets varies periodically on the order of 10 years, as shown in Figure~\ref{jupiter}.
Therefore, they cannot be used if the lowest elevation of the telescope is above them.
This difficulty was encountered by our telescope, GroundBIRD~\cite{GB} from 2019 to 2023 because the lowest elevation of the telescope is $60^{\circ}$. 
Therefore, a calibration source other than Jupiter or Saturn was required.
Although Mars and Venus are other candidates for pointing calibration, it is difficult to use them because they are located close to the Sun.
We excluded the observation region at a distance of $\pm 11^{\circ}$ from the Sun from the perspective of the GroundBIRD, which we describe in Section 2.
In addition, these planets (i.e., Jupiter, Saturn, Mars and Venus) are not sufficiently bright for detection at high signal-to-noise ratios because of the beam width of GroundBIRD (i.e., $\ang{0.6}$).

The Moon is a potential astronomical source for pointing calibration.
GroundBIRD can frequently observe the Moon at least once a month.
Its visible size is $\ang{;30}$ (i.e., $0.5^{\circ}$), and its brightness temperature is high ($\sim$ 200\,K).
Although ideally point sources are used for pointing calibrations, the Moon is sufficiently bright for measurements with a high signal-to-noise ratio.
Even though we used Jupiter which is the brightest of the point sources, telescopes whose beam width is sub-degrees need to integrate data to identify a position of it. Therefore, the measurement with the high signal-to-noise ratio introduces an advantage to reduce the effect of atmospheric fluctuation.
Recently, in the CLASS experiment, the Moon was used as a pointing calibration source because the beam width was sufficiently large ($1.5^{\circ}$) compared with the visible size of the Moon~\cite{CLASS}.
The beam width of GroundBIRD ($\ang{;36}$) is slightly larger than the Moon’s visible size, and the dynamic range of the detector response is sufficiently large to measure the Moon signal.
Therefore, Moon observational data can be used for pointing calibration, as demonstrated in the present study.

The methodology of the pointing calibration using the Moon has not been well established yet.
For instance, there has not been any systematic error study in 145 GHz.
In this study, we demonstrate the pointing calibration using the Moon. We also performed a systematic error study.
Because the Moon is brighter compared with the point-like planets, established methods in this study should be useful for other experiments whose beam width is sub-degrees as GroundBIRD.
In Section 2, the GroundBIRD telescope used in this study is described.
In Section 3, we describe the Moon’s angular response.
In Section 4, the methodology used to extract the central position of the Moon using each detector is described.
In Section 5, we define the pointing calibration model and present the calibration results.
In Section 6, we discuss the systematic uncertainties.
Finally, in Section 7, our conclusions are presented.

\begin{figure}[t!]
  \centering
  \includegraphics[width = 9cm]{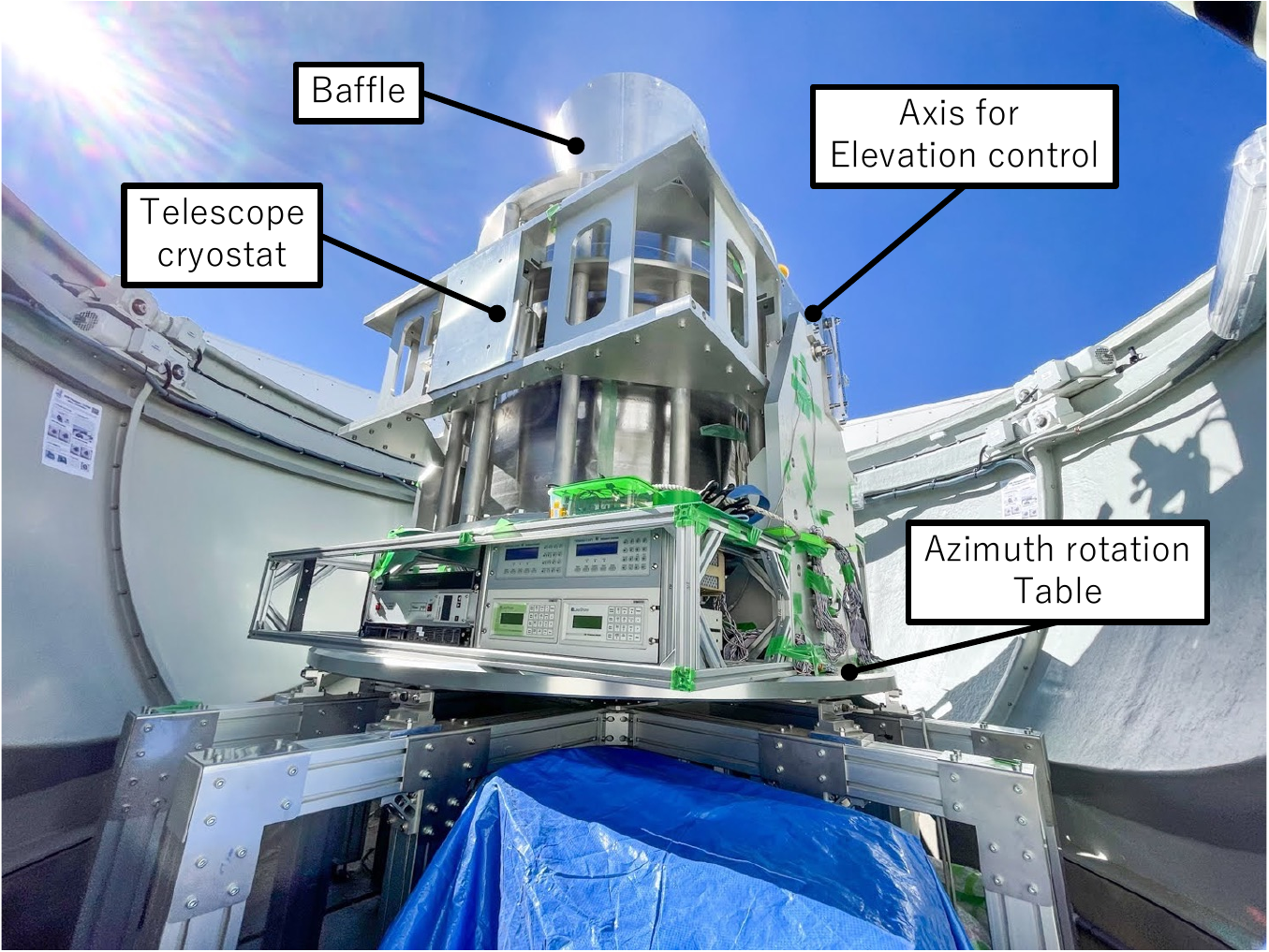}
  \caption{GroundBIRD telescope.
  The telescope cryostat is placed on the azimuth rotation table, and it is rotated in the azimuth with a fixed elevation.}
  \label{GB_pic}
  \centering
  \includegraphics[width = 0.8\linewidth]{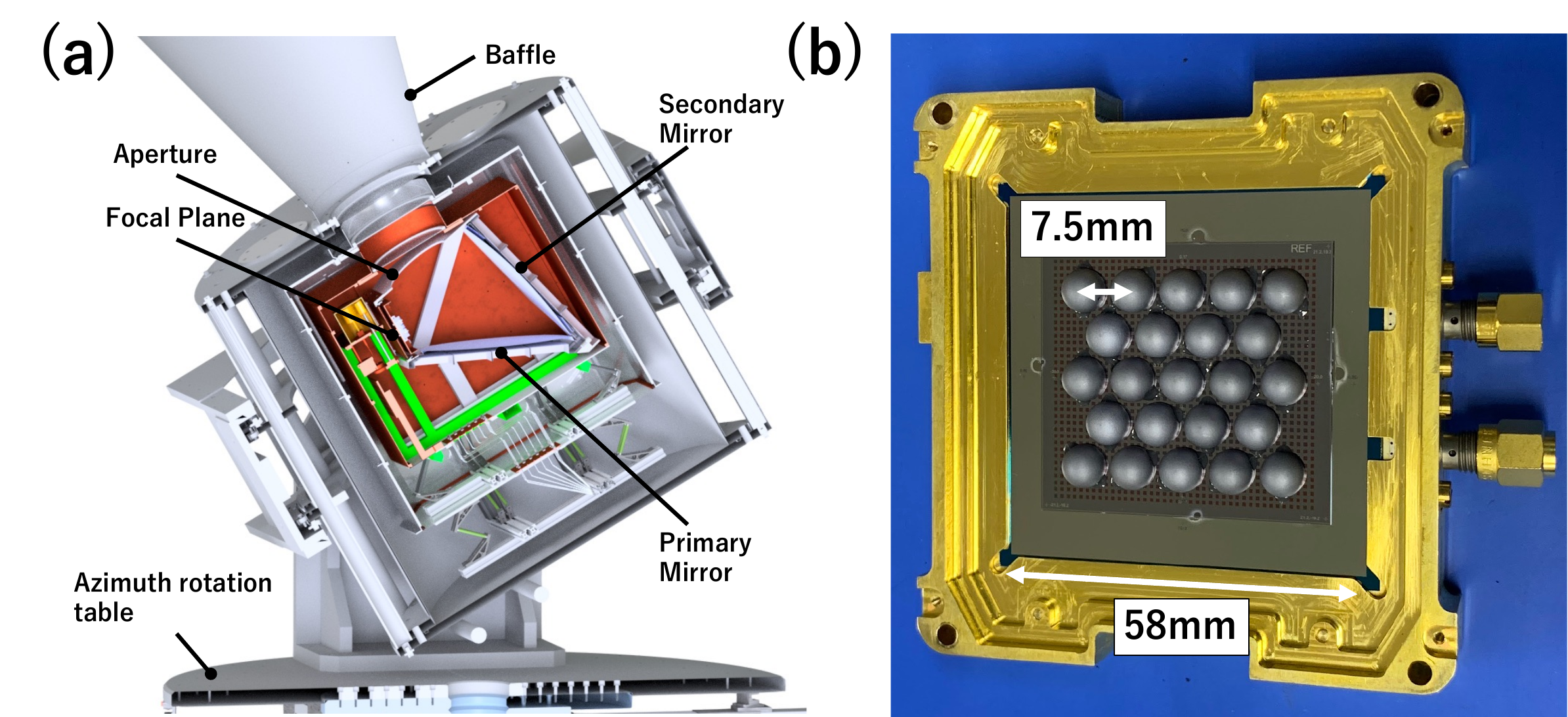}
  \caption{(a) Cross-sectional illustration of the GroundBIRD telescope. (b) Photograph of the detector array module used in this research. There are 23-antenna-coupled KIDs behind each silicon lens.}
  \label{GB_LT221}
\end{figure}

\section{GroundBIRD telescope}

GroundBIRD is a ground-based CMB polarization experiment (Figure~\ref{GB_pic}).
Our telescope is located at the Teide Observatory in the Canary Islands, Spain.
Its longitude, latitude, and altitude are $28^{\circ} 18^{'}$~N, $16^{\circ} 30^{'}$~W, and 2,400~m, respectively.
The telescope rotates continuously in the azimuthal direction at a fixed elevation.
The maximum speed of the azimuth rotation is 20 revolutions per minute (RPM).
This rapid scan modulation mitigates the effects of atmospheric fluctuations ~\cite{GB}.
The region above an elevation of $ 60^{\circ}$ can be observed, which allows for the observation of up to 45\% of the entire sky (declination range of $\ang{-2}$ -- $\ang{58}$).

Figure~\ref{GB_LT221} (a) shows a cross-sectional illustration of the GroundBIRD telescope.
The sky signal enters the cryostat through a polyethylene window~\cite{Komine} and is focused onto focal-plane detectors using a Mizuguchi–Dragone dual reflector~\cite{mirror1, mirror2, mirror3}.
The signals were detected by antenna-coupled Kinetic Inductance Detectors (KIDs)~\cite{MKID} using silicon lenses in front of each antenna.
The responses of each KID were measured as variations of the resonant phase and frequency~\cite{GB_readout, ishitsuka}.
The data used in this study were obtained with a 1 kHz sampling rate using a 23-KID array module for the 145 GHz band, as shown in Figure~\ref{GB_LT221} (b)~\cite{hybrid_MKID, antenna}.\footnote{The module for the 145GHz band was only installed for the commissioning observations in 2021--2022. We fully upgraded it in May 2023. They were fabricated at the Delft University of Technology and the Netherlands Institute for Space Research in the Netherlands.}
They are maintained at approximately 280 mK.

\begin{figure}[t!]
  \centering
  \includegraphics[width = 6cm]{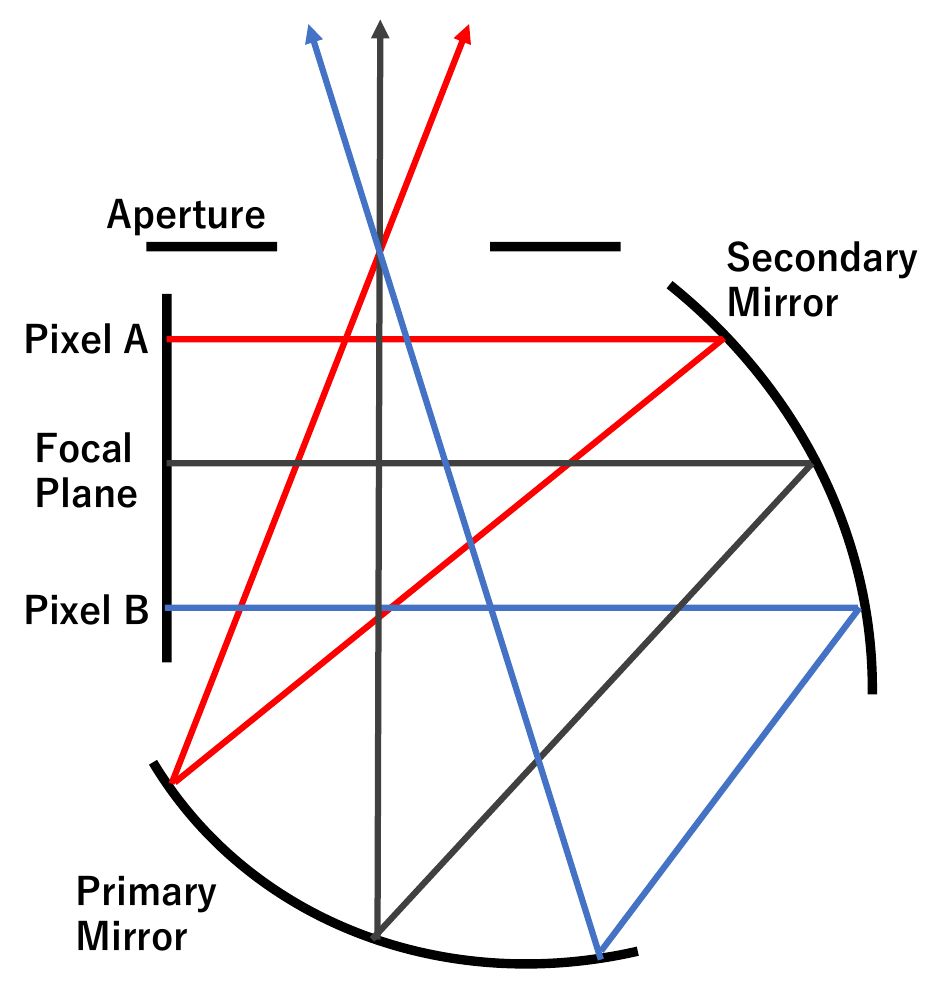}
  \caption{The pointing of each detector pixel is dependent on its location in the focal plane.
  The aperture diameter is 220 mm. 
  We use Mizuguchi–Dragone dual reflector which is a combination of a primary paraboloid and a secondary hyperboloid mirror.
  For two mirrors, diameters at optical area are the same. 
  The diameters for the minor-axis and the major-axis are 360 mm and 490 mm, respectively.
  }
  \label{collimation}
\end{figure}

According to the simulation study by the CST Microwave Studio, the beam width of the telescope is $\ang{;36}$ (i.e., $0.6^{\circ}$)~\cite{Jihoon}. Its ellipticity is at most 1\% according to this simulation.
\footnote{ In the reference~\cite{Jihoon}, we confirmed this simulation by the laboratory measurements with a small aperture setup ($\sim 1/4$ size) in the different frequency (90~GHz). We did not have sufficient space in the laboratory. We did not have a setup for 145~GHz with sufficient signal-to-noise ratio. These are the reasons for the small aperture and the different frequency. Validation using other planets such as Jupiter is a future study.}
As shown in Figure~\ref{collimation}, each detector in the focal plane points to a different position in the sky.
The angular interval between the collimations of each detector in the sky is typically $\ang{;51}$.
Therefore, the field of view of the telescope is $\pm 11^{\circ}$.
The elevation and azimuth of the telescope are monitored using a laser rotary encoder (R-1SL, Canon) and a magnetic rotary encoder (ERM220, HEIDENHAIN), respectively.
Their precisions are $\ang{;;4.0}$ and $\ang{;;3.4}$ for the elevation and azimuth, respectively~\cite{ikemitsu}.
Table~\ref{optics} lists the optical specifications of the GroundBIRD telescope.

In other CMB experiments, the ratio of pointing accuracy to their beam width is 0.13--0.14~\cite{PB,QUIET_point}.
In this case, the impact to the $B$-modes power measurement was estimated to be $\sim 10^{-4} \mu {\rm K^2}$ at multipole of 100, which approximately corresponds to the power at the tensor-to-scalar ratio ($r$) of 0.001~\cite{QUIET_Qband}. This is two orders of magnitude lower than our sensitivity with three years of observations ($r \approx 0.1$).
Based on this knowledge, we set the required pointing accuracy to $\ang{;4.7}$ (i.e., $\ang{;36} \times 0.13$) in this study.

\begin{table}[t!]
  \caption{Optical specifications of the GroundBIRD telescope.}
  \label{optics}
  \centering
  \scalebox{1}{ 
  \begin{tabular}{l r}
  \hline \hline
  Field of view                                              & $\pm 11^{\circ}$ \\
  Typical collimation interval between each detector pixel   & $\ang{;51}$ \\
  Beam width                                                 & $\ang{;36}$ \\
  Beam ellipticity                                           & $< 1\%$ \\
  Precision of elevation encoder                             & $\ang{;;4.0}$ \\
  Precision of azimuth encoder                               & $\ang{;;3.4}$ \\
  \hline
  Pointing accuracy (requirement)                            & $\ang{;4.7}$ \\
  \hline
  \label{optics}
  \end{tabular}}
\end{table}

\section{Angular response to the Moon}

\begin{figure}[tb!]
  \centering
  \includegraphics[width = 10cm]{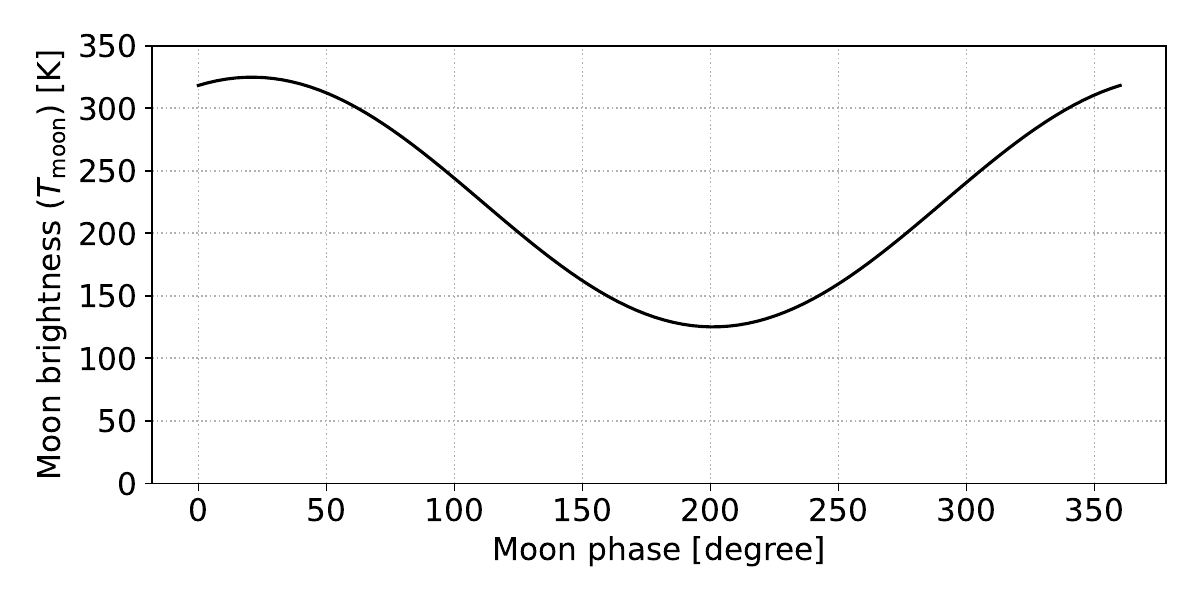}
  \caption{Brightness of the Moon signal as a function of the Moon phase for 145 GHz. The Moon phase at $\ang{0}$ corresponds to the full Moon.}
  \label{moonT}
\end{figure}

In the millimeter wavelength range, thermal radiation from the Moon’s surface is dominant compared with reflected sunlight.
The brightness of the Moon signal in Kelvin ($T_{\rm moon}$) is modeled using the Moon phase ($\psi$), as follows~\cite{T_model}:

\begin{align}
    &T_{\rm moon} = 225 \left\{ 1 + \frac{0.77}{\sqrt{1 + 2 \delta + 2 \delta^{2}}} \cos \left( \psi - \arctan \frac{\delta}{1+\delta } \right) \right\}, \label{eq:surfaceT}\\
        &\delta  \equiv 0.3 \lambda, \notag
\end{align}
where $\lambda$ is the wavelength in millimeters.
Figure~\ref{moonT} shows the brightness temperature at the Moon’s surface at 145 GHz ($\lambda \sim$ 2 mm) as a function of the Moon’s phase.
The timing of the maximum brightness is delayed compared with the full Moon owing to a delay in the temperature variation at the Moon’s surface.
The maximum and minimum Moon brightness temperatures were 325 and 125\,K, respectively.

\begin{figure}[t!]
  \centering
  \includegraphics[width = \linewidth]{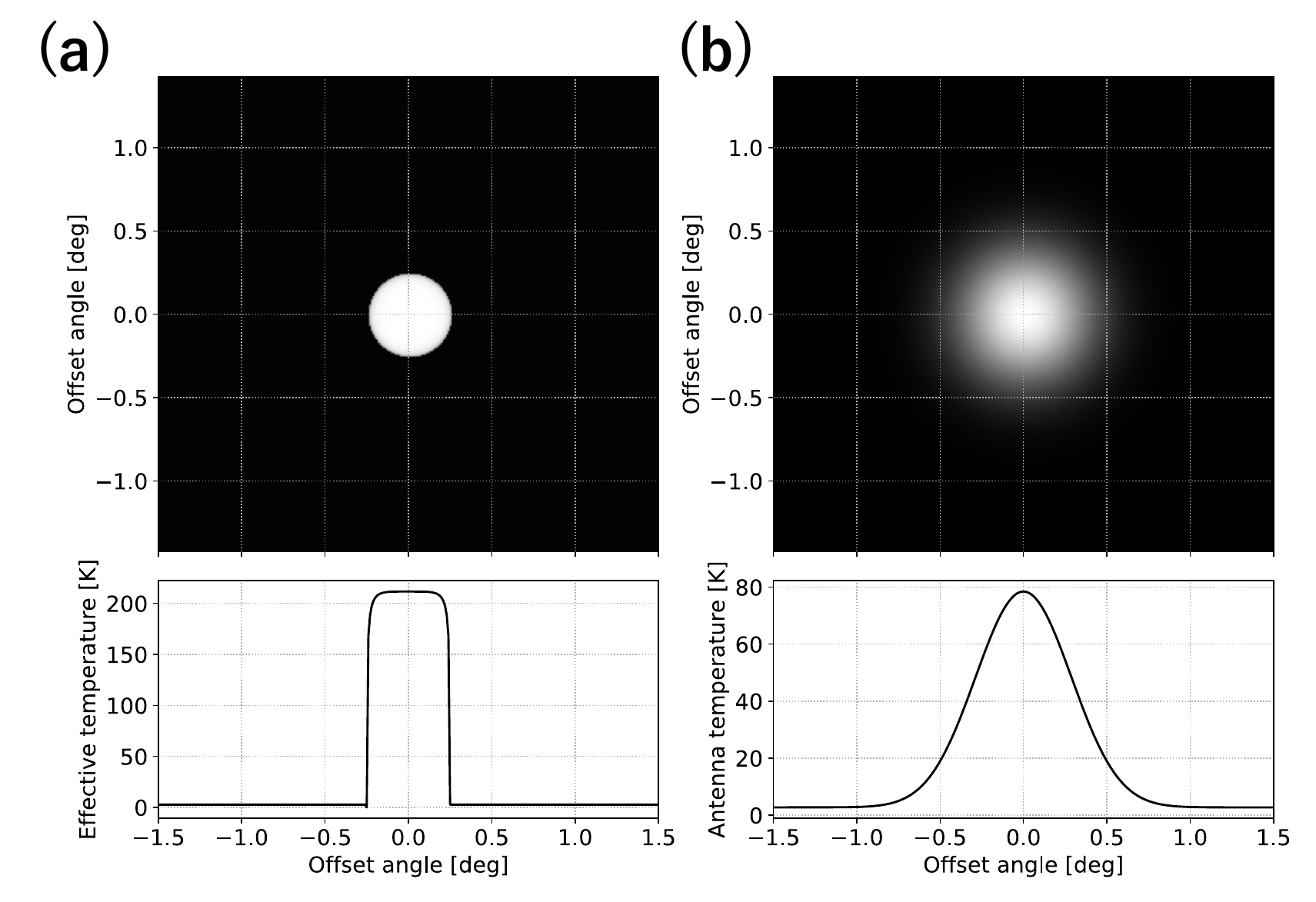}
  \caption{(a) Distribution of visible brightness temperature around the Moon from the Earth.
  (b) Convolved distribution with the beam modeled using the simple Gaussian (beam width of $\ang{;36}$).
  The plots in the bottom row present a cross-sectional view.}
  \label{moonmodel}
\end{figure}

From the Earth, the visible brightness ($T_{\rm eff}$) at the edge of the Moon is different from $T_{\rm moon}$ because of the different refractive indices of the Moon's surface.
Considering the Moon as a black body sphere covered by dielectric material layer~\cite{moon_model}, we obtained the angular distribution of the visible brightness, as shown in Figure~\ref{moonmodel} (left).
This is almost a top-hat distribution with a diameter of $\ang{;30}$.
The mean of $T_{\rm eff}$ within the visible range of the Moon is equal to the brightness temperature ($T_{\rm moon}$).
Outside the Moon, $T_{\rm eff}$ corresponded to the CMB temperature (2.725 K).
For real observations, the measured signals were convolved with the beam.
We used an angular response model convolved with $T_{\rm eff}$ and a Gaussian beam, as shown in Figure~\ref{moonmodel} (right).
The uncertainties owing to the nonuniformity of $T_{\rm moon}$ and the beam shape are discussed in Section 6.

\section{Reconstruction of the Moon's position}

We performed 19 observations of the Moon using an azimuth rotation scan at 10 RPM with a fixed elevation at $\ang{70}$ between the $8^{th}$ of February 2022 and the $18^{th}$ of April 2022, as listed in Table~\ref{observation}.
The rotation of the Earth then leads the Moon to form a map through telescope scans.
With regards to the Moon observations, we used time-ordered data (TOD) for each detector as well as telescope encoder data.
The duration of each observation was 38–60 min.
One of the 23 detectors was not used because its resonant frequency was outside our readout bandwidth.
Prior to the analysis, we made corrections for the microwave phase delay owing to the cable length and detector responses~\cite{Kutsuma1}.
\begin{table}[t!]
  \caption{Information on the Moon observations used in this study.}
  \label{observation}
  \centering
  \scalebox{0.9}{ 
  \begin{tabular}{c c c c c}
  \hline \hline
  Observation date(UTC)  & Duration [min] &  Ascent or decent   & Azimuth [deg]  & Elevation [deg] \\
  \hline
  ~8/2/2022 17:11 - 18:11  &  60  &  ascent & 106 - 126  & 61 - 73 \\
  ~8/2/2022 20:14 - 21:01  &  47  &  decent & 238 - 253  & 63 - 72 \\
  14/2/2022 22:12 - 22:50  &  38  &  ascent & ~95 - 103  & 64 - 72 \\
  15/2/2022 22:56 - 23:56  &  60  &  ascent & 101 - 120  & 62 - 74 \\
  16/2/2022 23:48 - 24:48  &  60  &  ascent & 110 - 133  & 61 - 72 \\
  17/2/2022 02:36 - 03:36  &  60  &  decent & 224 - 247  & 61 - 72 \\
  ~7/3/2022 15:29 - 16:29  &  60  &  ascent & 115 - 146  & 65 - 75 \\
  ~7/3/2022 18:01 - 19:01  &  60  &  decent & 230 - 252  & 61 - 72 \\
  ~8/3/2022 15:47 - 16:47  &  60  &  ascent & 100 - 117  & 62 - 74 \\
  ~9/3/2022 16:30 - 17:30  &  60  &  ascent & ~93 - 104  & 62 - 75 \\
  ~9/3/2022 19:50 - 20:43  &  52  &  decent & 258 - 268  & 62 - 73 \\
  10/3/2022 20:42 - 21:36  &  54  &  decent & 266 - 273  & 61 - 73 \\
  ~5/4/2022 17:41 - 18:28  &  47  &  decent & 255 - 265  & 63 - 73 \\
  ~6/4/2022 15:26 - 16:26  &  60  &  ascent & ~91 - 103  & 65 - 78 \\
  ~6/4/2022 18:34 - 19:26  &  53  &  decent & 264 - 272  & 62 - 73 \\
  ~7/4/2022 19:25 - 20:23  &  58  &  decent & 269 - 275  & 61 - 73 \\
  ~8/4/2022 16:50 - 17:50  &  60  &  ascent & ~86 - ~92  & 62 - 75 \\
  ~8/4/2022 20:16 - 21:16  &  60  &  decent & 269 - 275  & 60 - 73 \\
  12/4/2022 20:28 - 21:28  &  60  &  ascent & 114 - 140  & 62 - 73 \\
  \hline
  \end{tabular}
  }
\end{table}
\begin{figure}[t!]
  \centering
  \includegraphics[width = \linewidth]{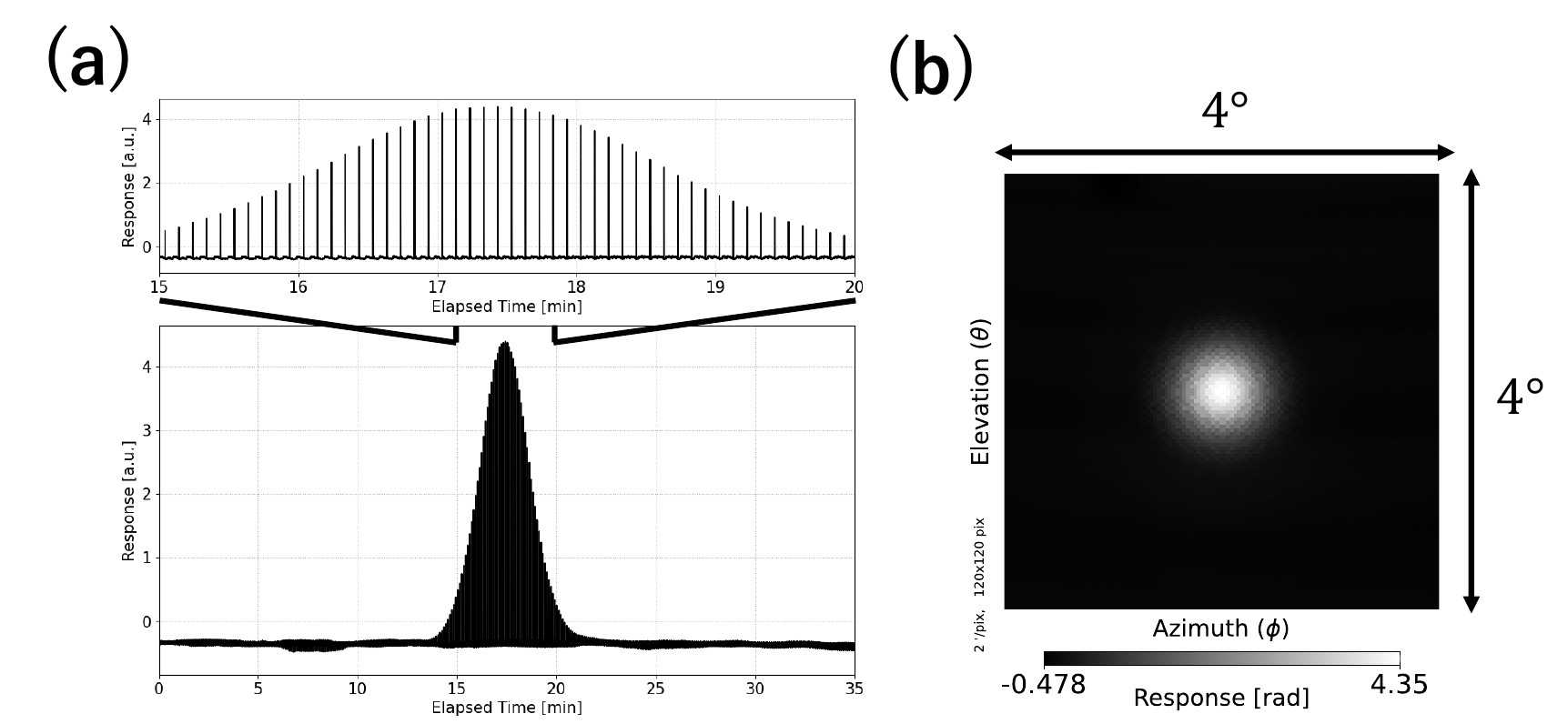}
  \caption{(a) Detector response as a function of the elapsed time.
  (b) Reconstructed Moon image using the Moon-centerd coordinates for azimuth (horizontal axis) and elevation (vertical axis).
  This image is created using \textsf{healpy} with the parameter $N_{\rm side} = 1024$.
  The baseline fluctuations due to atmospheric radiation are subtracted as described in the text.}
  \label{tod_mcmap}
\end{figure}
\begin{figure}[tb!]
  \centering
  \includegraphics[width = \linewidth]{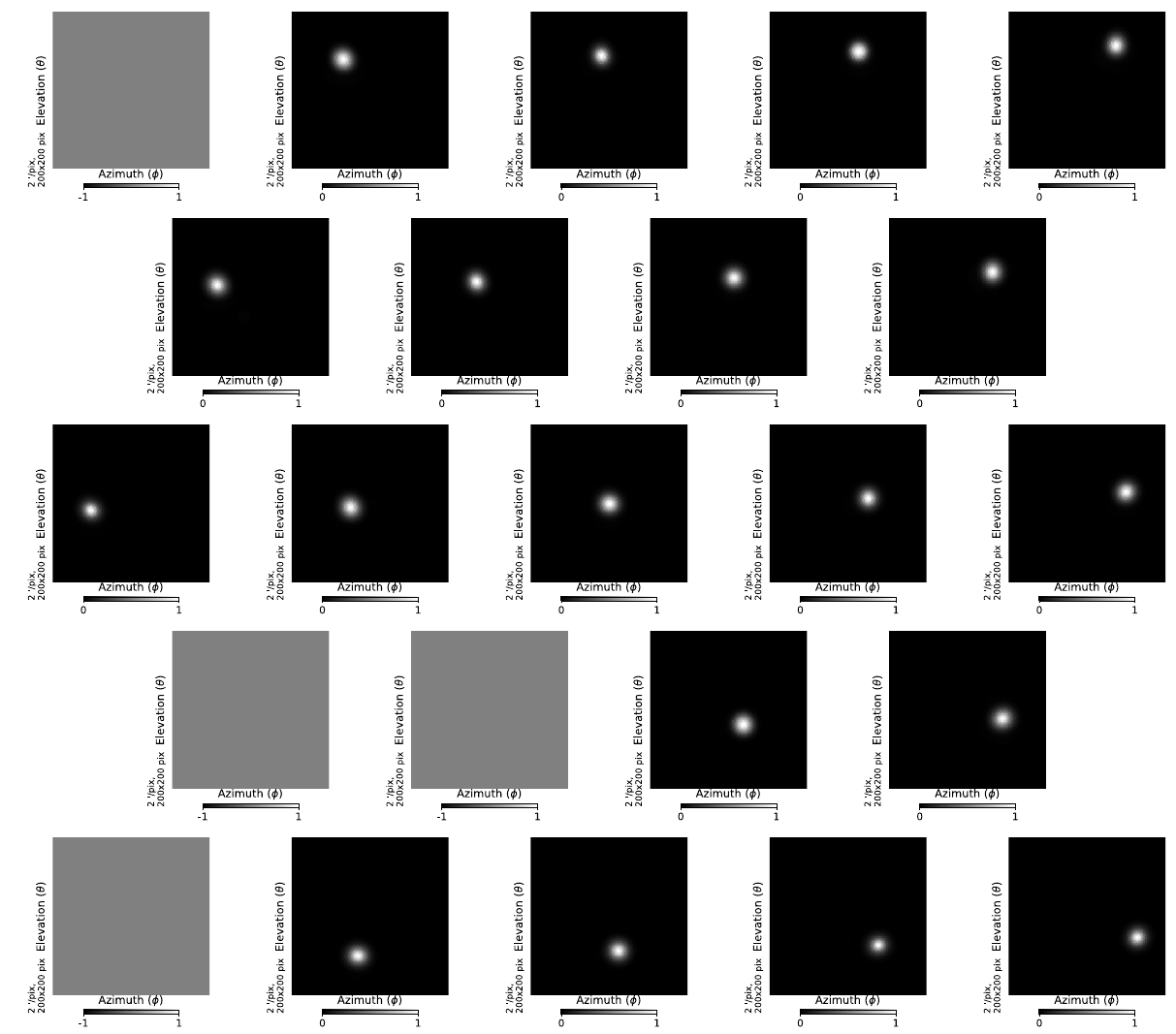}
  \caption{Reconstructed Moon images for each detector in the Moon-centered coordinates.
  The highest signal is normalized to 1.
  The positions of the plots indicate the locations of each detector pixel on the focal plane.
  The data of the four detectors are not used; they are indicated by the gray squares.
  The locations of the Moon images in each detector image are different because each detector has a different collimation offset.
  The angular ranges in all plots are $\ang{;400}$ times $\ang{;400}$.
  These images are created using \textsf{healpy} with the parameter $N_{\rm side} = 1024$.}
  \label{moon_each}
\end{figure}
We selected good-quality data from each detector for each observation based on the following criteria:
\begin{itemize}
    \item The intervals of the resonant frequencies for each detector must be greater than 0.5 MHz for the same observation. 
    This condition eliminates potential crosstalk among the detectors.
    We did not use two of the 22 detectors due to this criterion.
    \item Data within an elevation of $1.5^{\circ}$ from the Moon center must be obtained continuously for each detector, i.e., each detector must observe the outside region of the Moon as well as the Moon itself.
    \item 
    The Moon signal must be within the dynamic range of each detector.
    The detector response is proportional to $\tan(\frac{\psi}{2})$ where $\psi$ is a phase of fed microwaves for the detector readout~\cite{gao}. Thus, we observe a rapid jump in the detector response when the signal intensity is higher than the dynamic range.
    We identified it as rapid variations of detector response (more than 30 times) within the visible size of the Moon in the azimuthal scan.
    We did not use one of the 20 detectors based on this criterion.
\end{itemize}
On the basis of these criteria, 345 Moon observation data were selected.

\begin{figure}[tb!]
  \centering
  \includegraphics[width = 0.7\linewidth]{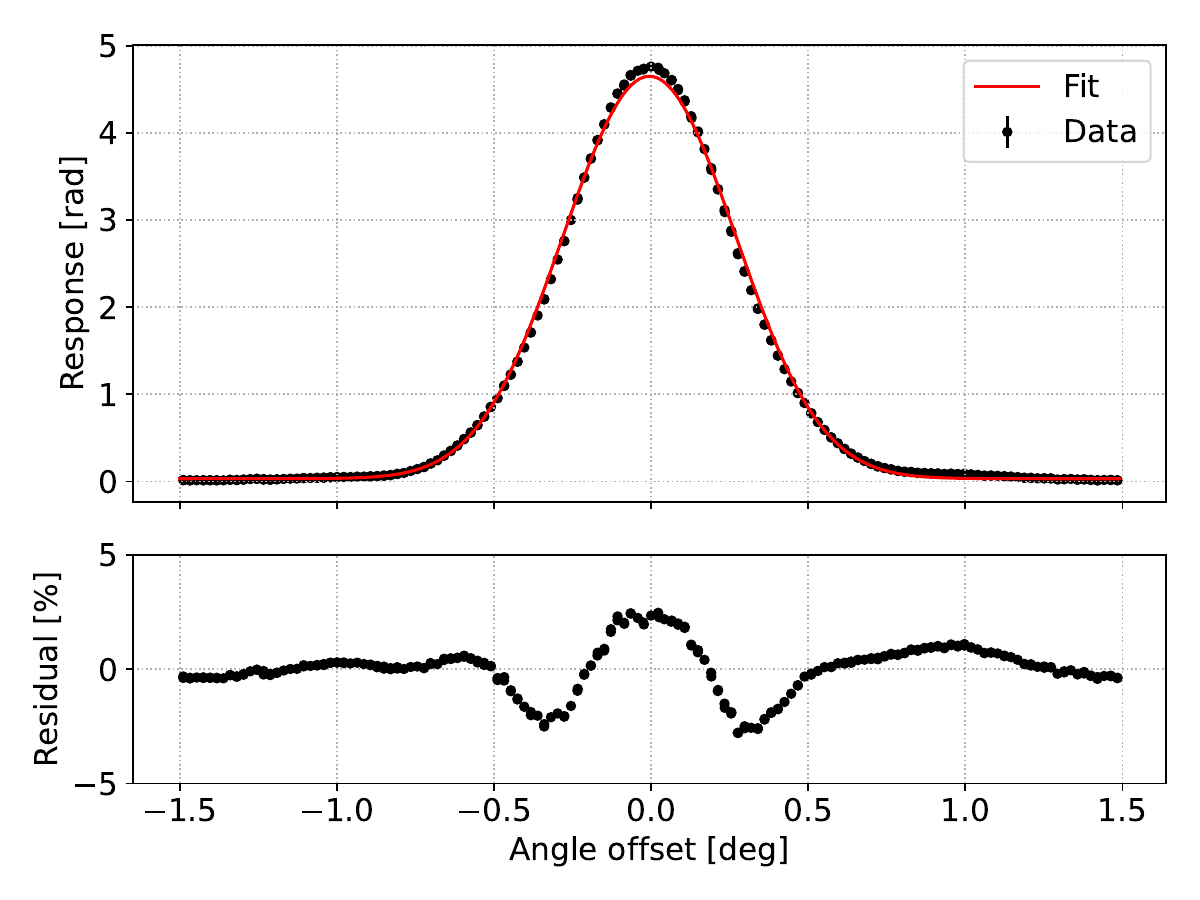}
  \caption{Data from the Moon scan fitted with a Gaussian beam in azimuthal scan. The non-Gaussianity of the beam shape leads to approximately 3\% residuals because we use the simple Gaussian model in the fit. This effects is considered to be a systematic error, as described in the text (Section 6).}
  \label{fit_sec}
\end{figure}

Figure~\ref{tod_mcmap} (a) presents the TOD for a single observation.
The overall shape of the bottom figure traces the elevation pattern of the Moon, and the spikes in the top figure trace the azimuth pattern of the Moon.
The time information between the detector data and telescope encoder data was synchronised by distributing a common clock signal.
Thus, a map of the elevation and azimuth coordinates for each detector was constructed.
For each TOD point, we can calculate the true position of the Moon using \textsf{astropy}.
Using the true position of the Moon, we further converted the map to Moon-centred coordinates, the axes of which comprise the elevation and azimuth from the Moon’s center.
One of the reconstructed Moon images in the Moon-centered coordinates is shown in Figure~\ref{tod_mcmap} (b).
In this study, we subtracted the baseline offset for each azimuthal scan to eliminate the effects of atmospheric fluctuations.
The baseline offset should be calculated from the detector response at away from the Moon. 
Therefore, each baseline offset was calculated as the mean value of the signals between $1.5^{\circ}$ and $2.0^{\circ}$ from the observed center of the Moon.
Figure~\ref{moon_each} presents the reconstructed moon images obtained from each detector as one of the observations.
The locations of the observed Moon images are different for each detector because each detector has a different collimation point in the sky, as described in Section 2.

\begin{figure}[t!]
  \centering
  \includegraphics[width = \linewidth]{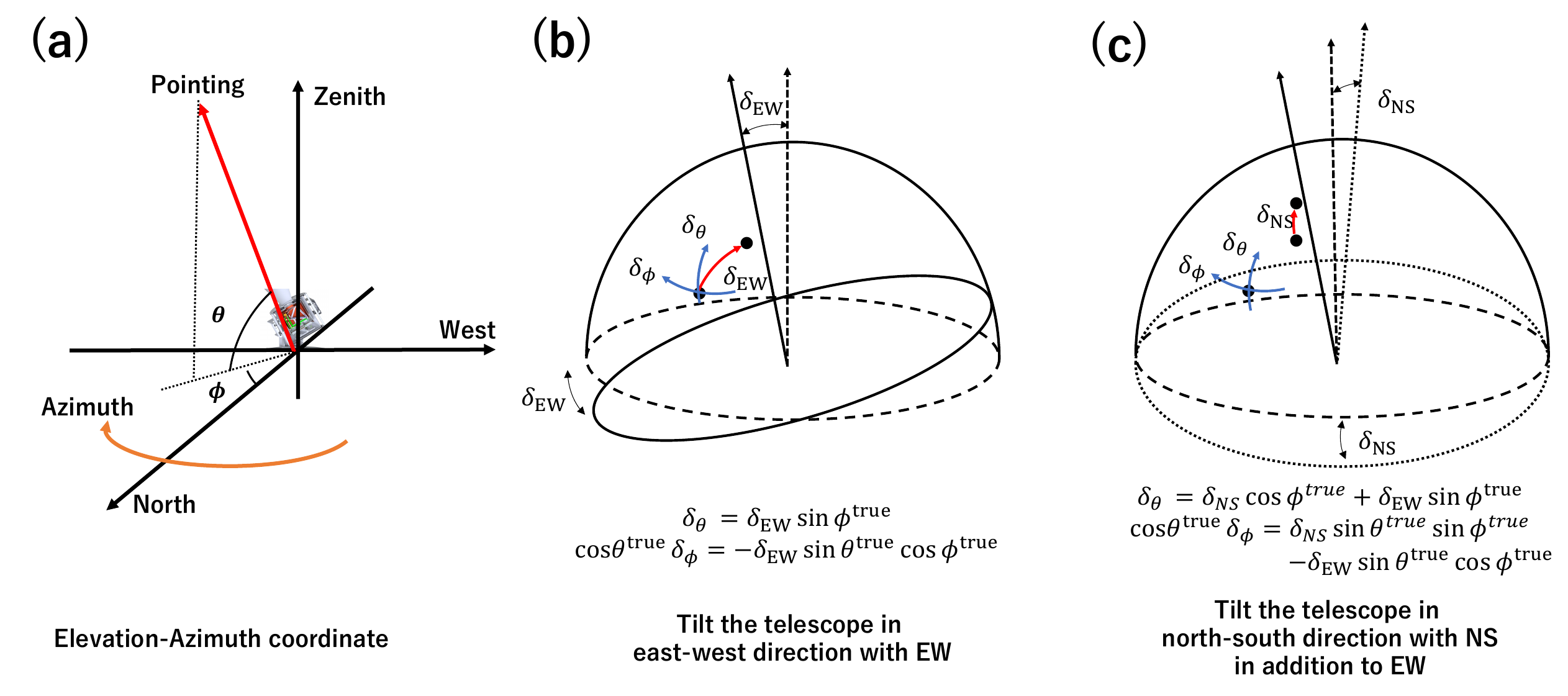}
  \caption{(a) Definition of the elevation ($\theta$) and azimuth ($\phi$).
  (b) Schematics of the axis tilt to the east--west direction.
  (c) Schematics of the axis tilt to the east--west direction and north--south direction.
  }
  \label{coordinate}
  \centering
  \includegraphics[width = \linewidth]{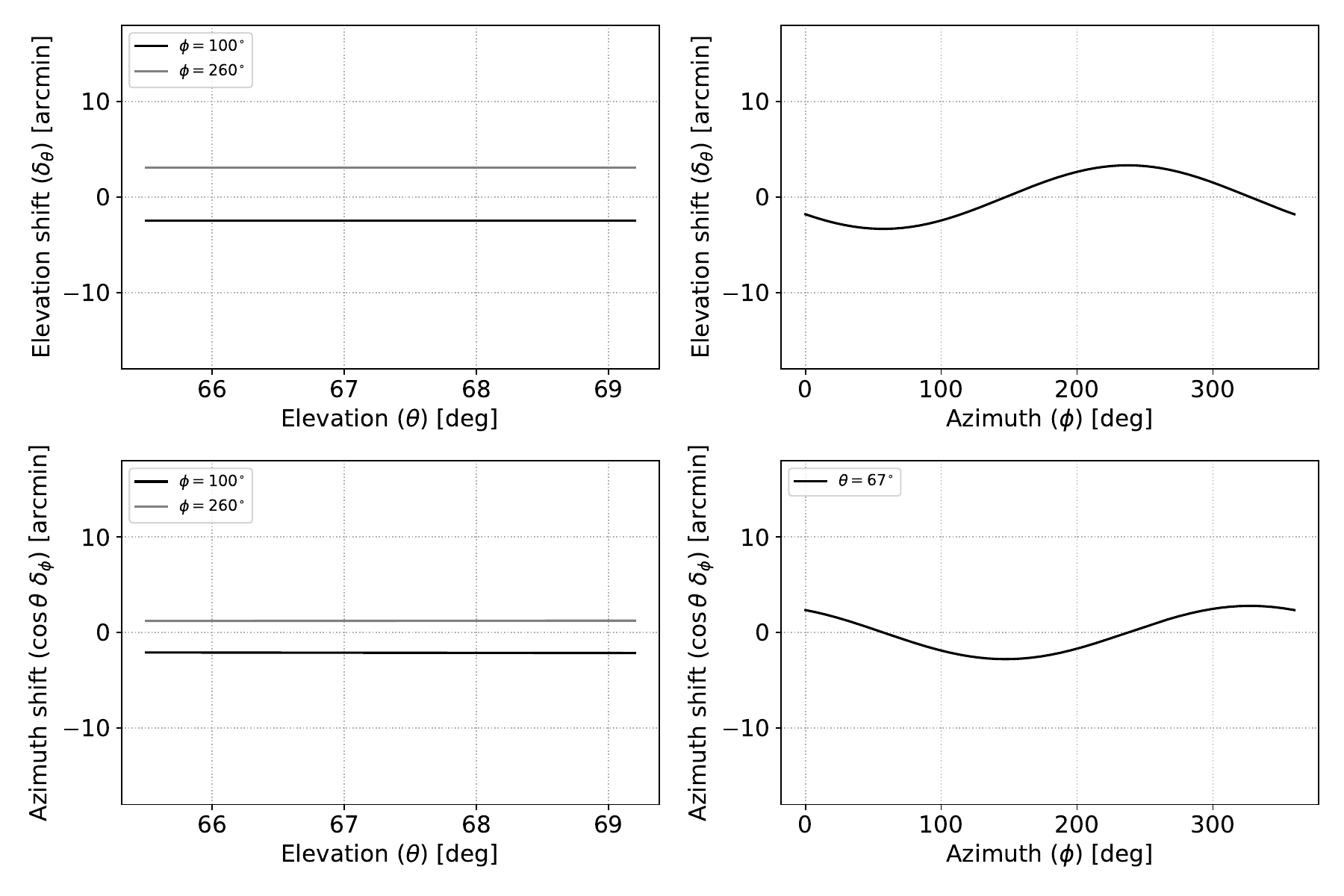}
  \vspace{-10mm}
  \caption{Pointing shifts as a function of the elevation or azimuth with $\delta_{\rm NS}$ of $-\ang{;1.8}$ and $\delta_{\rm EW}$ of $-\ang{;2.8}$.
  The shift in the elevation depends on the azimuth pointing.
  In contrast, the shift in the azimuth depends on both the elevation and the azimuth pointing.}
  \label{delAE}
\end{figure}

Based on the Moon’s response defined in Section 3, we extracted the Moon’s center positions for each detector and for each observation, by performing an unbinned likelihood fit for the data within a radius of $1.5^{\circ}$ from the Moon’s center.
The fitting parameters were the elevation and the azimuth of the Moon center, detector gain, beam width, and constant offset as the background residual.
Figure~\ref{fit_sec} presents an example of a result obtained by the fitting procedure.
The difference in the beam between the data and the model has a negligible effect on the pointing calibration, which is discussed in Section 6.
The extracted Moon positions for each detector and each observation comprised 345 samples and were used to calibrate the pointing model described in the next section.
The average collimation interval (a distance between neighbor detectors at the sky) was $\ang{;52} \pm \ang{;1}$, which is consistent with the design ($\ang{;51}$).

\section{Pointing calibration}

The tilts of the elevation and azimuth axes produce pointing shifts in terms of the elevation ($\theta$) and azimuth ($\phi$), as shown in Figure~\ref{coordinate}.
Therefore, we modeled the pointing shifts as follows~\cite{pointing_model}:
\begin{subequations}
    \begin{empheq}[left = {\empheqlbrace \,}, right = {}]{align}
    \delta_{\theta} (\phi) &= \delta_{\rm NS} \cos{\phi} + \delta_{\rm EW} \sin{\phi}, \label{eq:pointing_delE_axis}\\
    \cos{\theta} \delta_{\phi} (\theta , \phi) &= \delta_{\rm NS} \sin{\theta} \sin{\phi} - \delta_{\rm EW} \sin{\theta} \cos{\phi}, \label{eq:pointing_delA_axis}
    \end{empheq}
\end{subequations}
where $\delta_{\rm NS}$ is the tilt in the north--south direction and $\delta_{\rm EW}$ is the tilt in the east--west direction.
Figure~\ref{delAE} shows pointing shifts as a functions of the true elevation and the true azimuth for $\delta_{\rm NS} = -\ang{;1.8}$ and $\delta_{\rm EW} = -\ang{;2.8}$ calculated from Eq. (\ref{eq:pointing_delE_axis}) and Eq. (\ref{eq:pointing_delA_axis}).

In addition to the tilts of the axes, the encoder and collimation offsets of each detector cause constant pointing shifts, as shown in Figure~\ref{moon_each}.
We define the parameters for the encoder offsets in terms of elevation ($\theta_e$) and azimuth ($\phi_e$).
We also define the collimation offsets of each detector for the elevation ($\theta_i$) and azimuth ($\phi_i$).
Here, $i$ is the index for each detector.
Using these parameters, the pointing model for the data is defined as follows:

\begin{subequations}
    \begin{empheq}[left = {\empheqlbrace \,}, right = {}]{align}
    \theta^{\rm model} &= \theta^{\rm data} - \theta_e - \theta_i - \delta_{\theta} (\phi^{\rm true}), \label{eq:pointing_delE_tilt_zero_coll} \\
    \phi^{\rm model} &= \phi^{\rm data} - \phi_e - \phi_i- \delta_{\phi} (\theta^{\rm true}, \phi^{\rm true}), \label{eq:pointing_delA_tilt_zero_coll}
    \end{empheq}
\end{subequations}
where $\theta^{\rm data}$ and $\phi^{\rm data}$ are the reconstructed Moon positions for each detector for each observation (i.e., the reconstructed data samples in Section 4), and $\theta^{\rm true}$ and $\phi^{\rm true}$ are the true positions of the Moon, which are calculated using \textsf{astropy}.

\begin{figure}[t!]
  \centering
  \includegraphics[width = 1.0\linewidth]{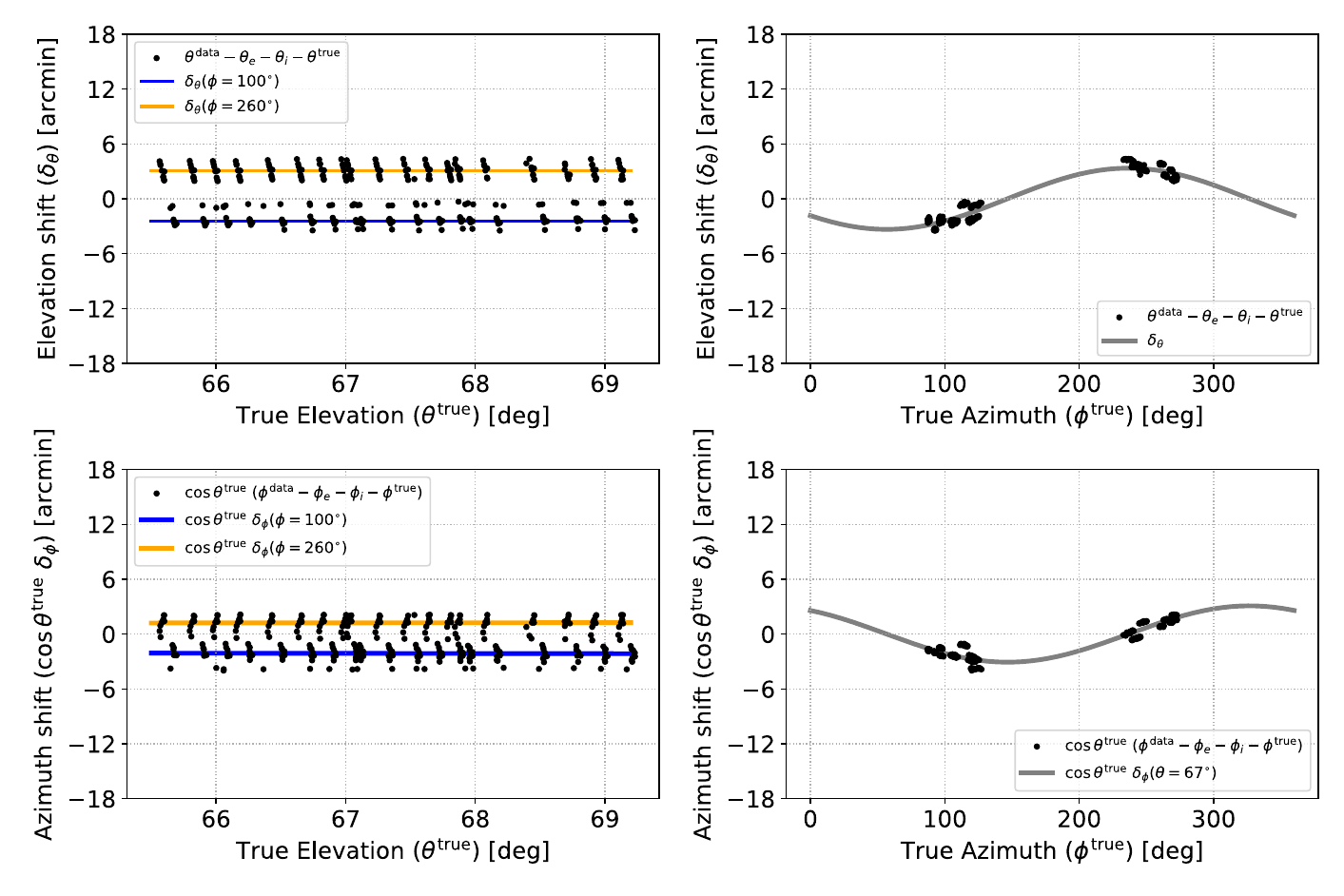}
  \vspace{-10mm}
  \caption{Pointing shifts as a function of the elevation or azimuth of the true Moon position.
  The points comprise reconstructed data calculated using Eq.~(\ref{eq:pointing_shiftE}) and Eq.~(\ref{eq:pointing_shiftA}).
  The lines were calculated using Eq.~(\ref{eq:pointing_delE_axis}) and Eq.~(\ref{eq:pointing_delA_axis}) with the same values for $\delta_{\rm NS}$ and $\delta_{\rm EW}$.}
  \label{result_diff}
\end{figure}

\begin{figure}[t!]
  \centering
  \includegraphics[width = \linewidth]{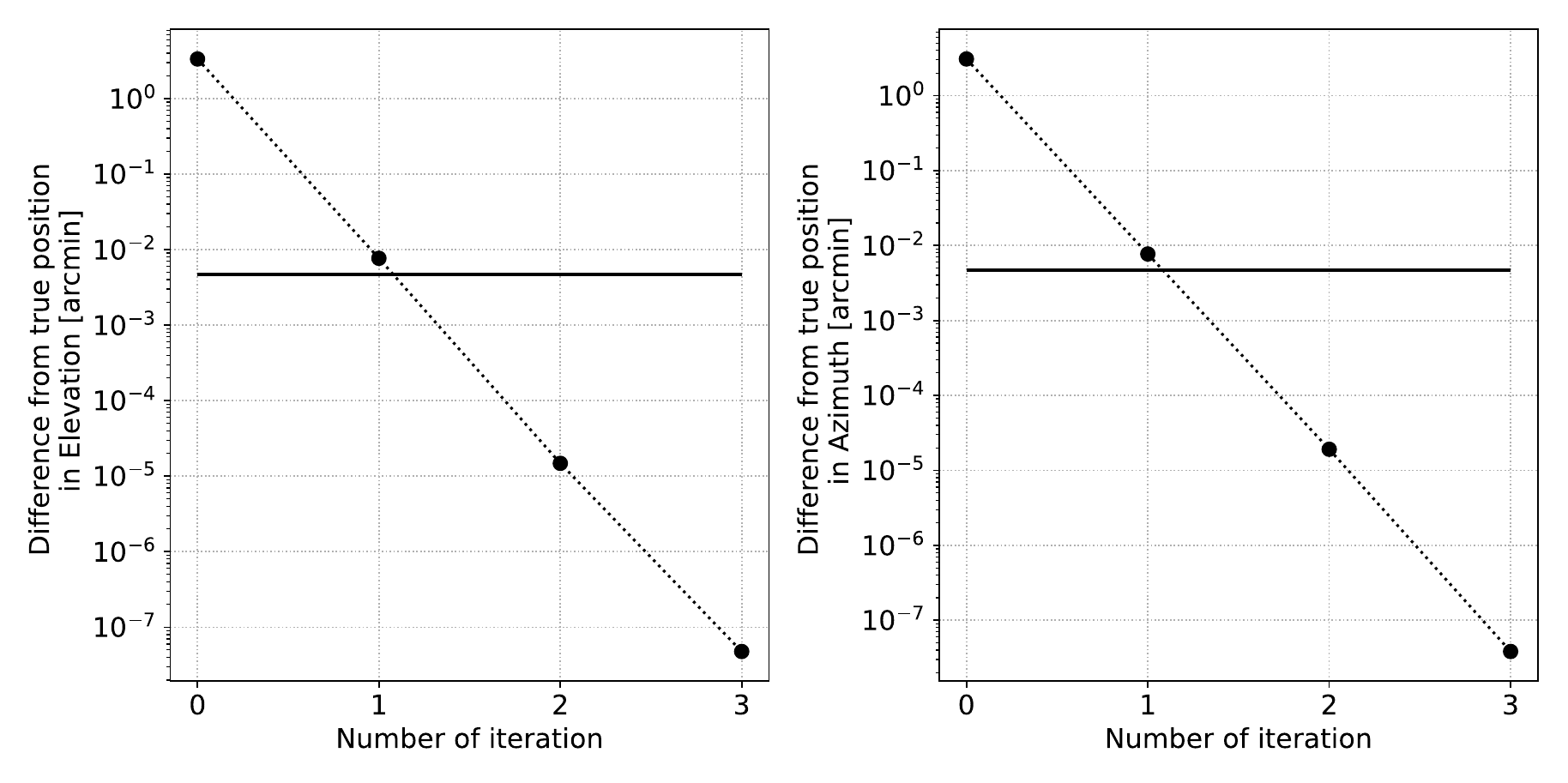}
  \caption{Pointing differences between a calibrated position with Eq.~(5a) -- Eq.~(6b) from the true position as a function of the number of iterations for the elevation and azimuth.
  The elevation angle is assumed to be $67^{\circ}$, which is a typical elevation angle.
  The solid lines are 1/1000 of the requirement.
  We realize a sufficient precision at $n \geqq 2$.
  }
  \label{ite_sim}
\end{figure}

These parameters were optimised to minimise the angular distance ($dl$) between the true and model positions; that is, $ dl^2 = (\theta^{\rm model} - \theta^{\rm true})^2 + [\cos{\theta^{\rm true}}(\phi^{\rm model} - \phi^{\rm true})]^2$.
Although $dl$ is calculated under the approximation of the small angular distance, this calculation method makes negligible effect ($\le \ang{;;0.01}$).
The optimized tilt angles were $\delta_{\rm NS} = \ang{;-1.8}$ and $\delta_{\rm EW} = \ang{;-2.8}$.
Using the extracted encoder and collimation offsets, we calculated the pointing shifts as follows:
\begin{subequations}
    \begin{empheq}[left = {\empheqlbrace \,}, right = {}]{align}
    \delta_{\theta} (\phi^{\rm true}) &= (\theta^{\rm data} - \theta_e - \theta_i) - \theta^{\rm true}, \label{eq:pointing_shiftE} \\
    \cos{\theta^{\rm true}} \delta_{\phi} (\theta^{\rm true}, \phi^{\rm true}) &= \cos{\theta^{\rm true}} [(\phi^{\rm data} - \phi_e - \phi_i) - \phi^{\rm true}]. \label{eq:pointing_shiftA}
    \end{empheq}
\end{subequations}
Figure~\ref{result_diff} presents the pointing shifts as functions of the true Moon position at that elevation or azimuth.
For comparison, we overlay the lines calculated using Eq. ~(\ref{eq:pointing_delE_axis}) and Eq.~(\ref{eq:pointing_delA_axis}).
These values were consistent within the required precision ($\ang{;4.7}$).

In the CMB analysis, we calculate the pointing based on the optimized model.
Although $\theta^{\rm true}$ and $\phi^{\rm true}$ are unknown in Eq.~(\ref{eq:pointing_delE_tilt_zero_coll}) and Eq.~(\ref{eq:pointing_delA_tilt_zero_coll}), accurate pointing can be calculated iteratively with sufficient precision~\cite{QUIJOTE_phD} instead of solving Eq.~(\ref{eq:pointing_delE_tilt_zero_coll}) and Eq.~(\ref{eq:pointing_delA_tilt_zero_coll}) exactly.
The following formulae are used:
\begin{subequations}
    \begin{empheq}[left = {\empheqlbrace \,}, right = {}]{align}
    \theta^{(0)}_i &= \theta^{\rm data} - \theta_e - \theta_i, \\
    \phi^{(0)}_i &= \phi^{\rm data} - \phi_e - \phi_i,
    \end{empheq}
\end{subequations}
\begin{subequations}
    \begin{empheq}[left = {\empheqlbrace \,}, right = {}]{align}
    \theta^{(n)}_i &= \theta^{\rm data} - \theta_e - \theta_i - \delta_{\theta} \left(\phi^{(n-1)}_i\right), \label{eq:pointing_iteE}\\
    \phi^{(n)}_i &= \phi^{\rm data} - \phi_e - \phi_i- \delta_{\phi} \left(\theta^{(n-1)}_i, \phi^{(n-1)}_i\right), \label{eq:pointing_iteA}\\
    n &= 1,2,3,...,\notag
    \end{empheq}
\end{subequations}
where $n$ indicates the $n$-th iteration.
Using Eq.~(\ref{eq:pointing_iteE}) and Eq.~(\ref{eq:pointing_iteA}) and the extracted parameters, we numerically calculated the differences from the true position for each $n$. We could achieve sufficient precision at $n \geqq 2$, as shown in Figure~\ref{ite_sim}.
We conclude that two iterations should be performed in the calculation of the pointing to reduce this effect on the pointing calibration to a negligible level.

\begin{figure}[H]
  \includegraphics[width = \linewidth]{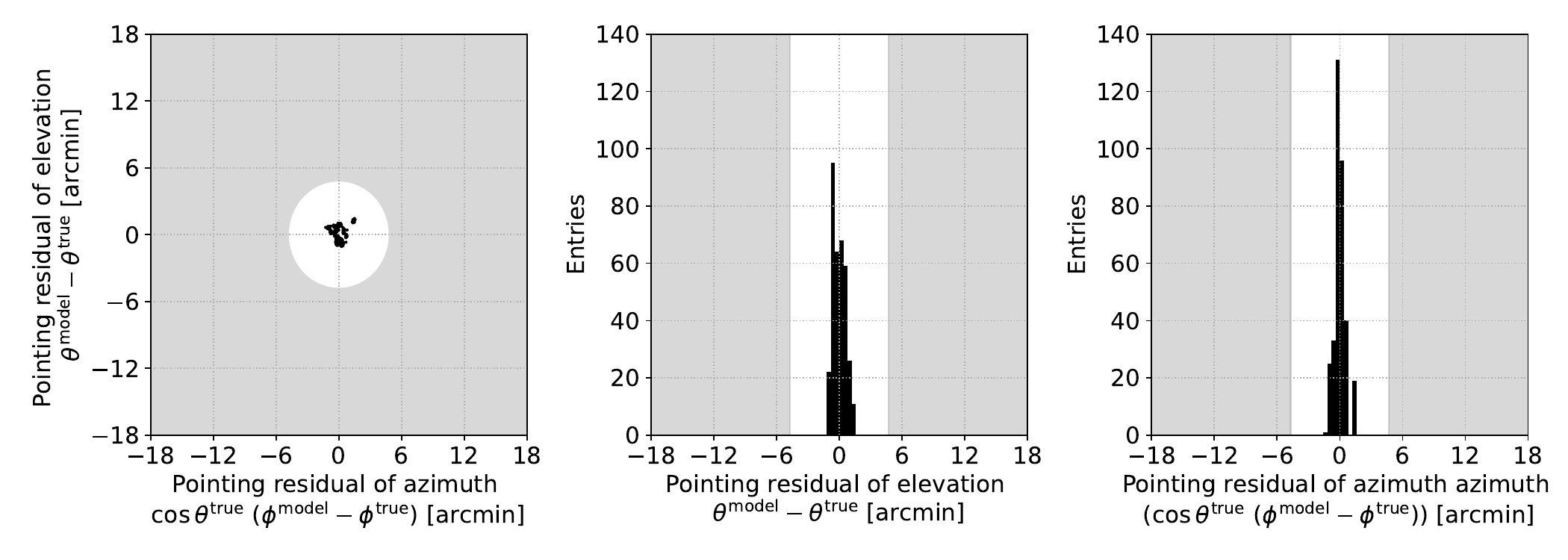}
  \caption{Residuals of the calibrated pointing for all 345 samples from each true position.
  The dynamic range of the elevation and azimuth axes corresponds to the beam width of the GroundBIRD ($\ang{;36}$).
  They are within the requirement, which is indicated by the unshaded regions.}
  \label{result_hist}
\end{figure}

\begin{figure}[t!]
  \centering
  \includegraphics[width = \linewidth]{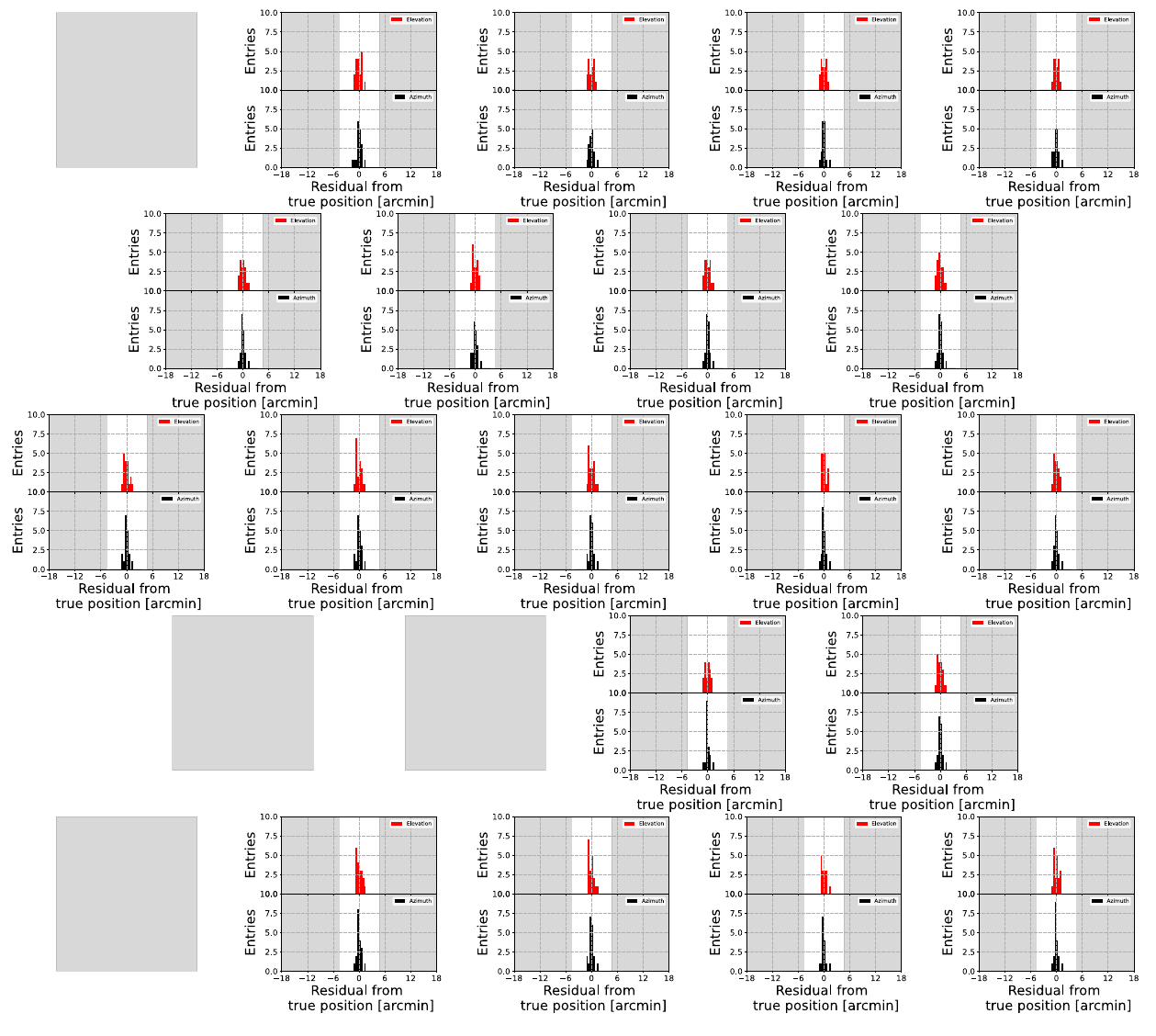}
  \caption{Pointing residuals of the elevation, $\theta^{\rm model} - \theta^{\rm true}$, and azimuth, $\cos{\theta^{\rm true}}~(\phi^{\rm model} - \phi^{\rm true})$, for each detector.
  Their mean values for each detector are consistent with zero.}
  \label{result_hist_each}
\end{figure}

Figure~\ref{result_hist} shows the residuals of the Moon position as reconstructed from the true positions for all 345 samples.
These values were within the range of our requirements.
Their root mean squares were $\ang{;0.6}$ for the elevation and $\ang{;0.5}$ for the azimuth times $\cos{\theta}$.
Similar plots for each detector are shown in Figure~\ref{result_hist_each}.
Their mean values were consistent with zero; that is, no bias was found.


\section{Systematic uncertainty}

The uncertainty associated with the beam shape is estimated by changing the response model of the fit.
We compared the results of the baseline analysis with a model that uses a fixed beam width as the design, one that uses an elliptical Gaussian beam with extracted ellipticity of $2.6 \pm 1.0 \%$ from the data, and one that uses a polynomial (i.e., non-linear) responsivity.
The pointing differences among the beam models are less than $\ang{;0.26}$. 
Note that the assumed ellipticity of the systematic error study is larger than that of the simulation study ($< 1\%$).
The uncertainties in the Moon’s position owing to the time constant of the detector and the  astronomical calculations are $\ang{;0.14}$ and $\ang{;0.31}$~\cite{astropy_error}, respectively.
The uncertainties in the elevation and azimuth encoders are $\ang{;0.066}$ and $\ang{;0.057}$, respectively, as described in Section 2.
Possible thermal effect due to ambient temperature change  is estimated by comparing two calibration results at high temperature (average of $10~^\circ $C) and low temperature (average of $3~^\circ $C).
We found the difference of $\ang{;0.4}$, while the statistical fluctuation is dominant in this comparison.
A possible mechanical variation associated with the scan is estimated from the difference between the residual from the pointing model at azimuth $< \ang{180}$ and that at azimuth $> \ang{180}$.
It was \ang{;0.057}.

\begin{table}[t!]
  \caption{Systematic uncertainties in pointing.}
  \label{systematics}
  \centering
  \scalebox{1}{ 
  \begin{tabular}{l r}
  \hline \hline
  Source                                   & [arcmin] \\
  \hline
  Beam shape                               & $2.6 \times 10^{-1}$ \\ 
  Moon position                            & $3.4 \times 10^{-1}$ \\ 
  Elevation encoder                        & $6.6 \times 10^{-2}$ \\
  Azimuth encoder                          & $5.7 \times 10^{-2}$ \\
  Ambient temperature                      & $4.0 \times 10^{-1}$ \\
  Mechanical variation                     & $5.7 \times 10^{-2}$ \\
  Non-uniformity temperature               & 3.2 \\
  \hline
  Total                                    & 3.2 \\
  \hline \hline
  \end{tabular}
  }
\end{table}

\begin{figure}[t]
  \centering
  \includegraphics[width = 15cm]{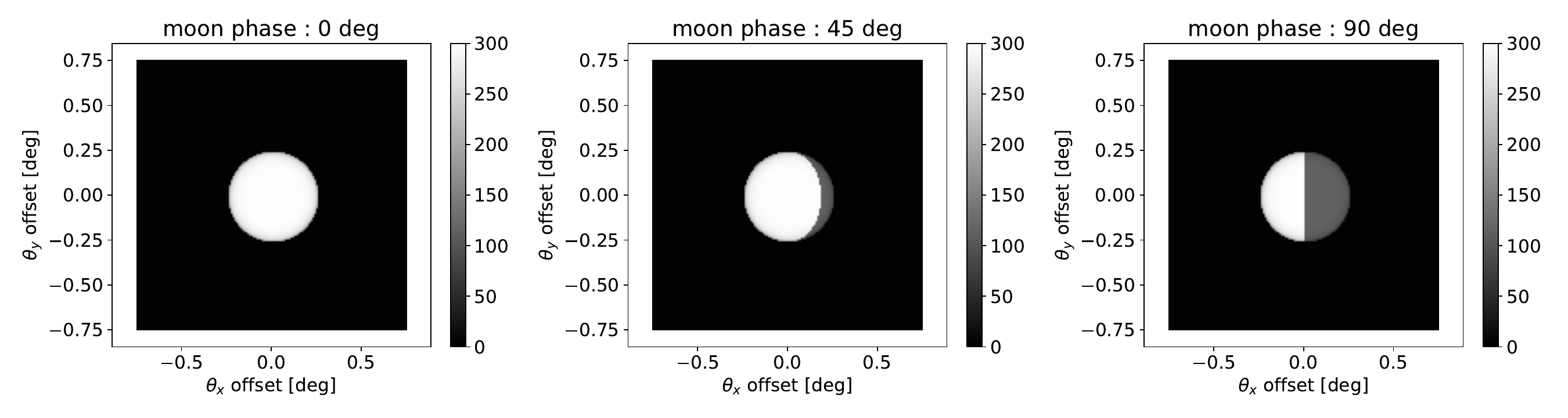}
  \caption{Distribution of the Moon brightness temperature, which varies with the Moon phase, where $\theta_{x}$ is the direction of the phase shift, and $\theta_{y}$ is perpendicular to it.
  These models are used only for the systematic error study.
  The white and gray regions correspond to the brightness temperatures of 325~K and 125~K, respectively.}
  \label{nonuni_model}
  \centering
  \includegraphics[width = 0.5\linewidth]{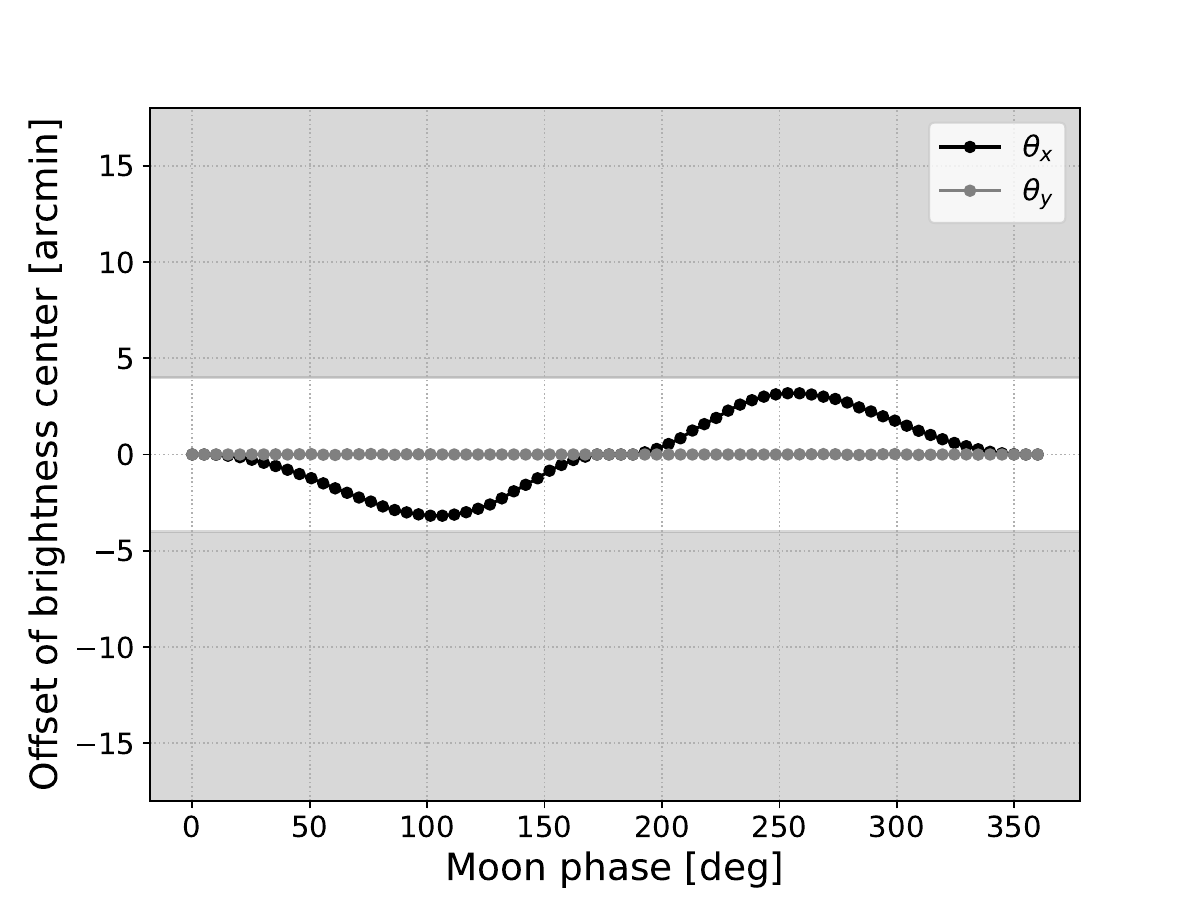}
  \caption{Offset of the brightness center as a function of the Moon phase, where $\theta_{x}$ and $\theta_{y}$ are the directions of the offsets, as shown in Figure~\ref{nonuni_model}.
  The maximum offset found here is smaller than the requirement, which is indicated by the unshaded area.}
  \label{center_diff}
\end{figure}

The systematic uncertainty is driven by the non-uniformity of the Moon’s brightness temperature.
We estimated this systematic uncertainty using a simulation based on simplified temperature distributions for each Moon phase, as shown in Figure~\ref{nonuni_model}.
Considering Eq.~(\ref{eq:surfaceT}), we set the maximum and minimum temperatures in the bright and shaded regions, respectively.
This assumption yielded the most conservative systematic error.
Using the same method as that used for the real data analysis, we extracted the central position of the Moon for the simulation data.
Figure~\ref{center_diff} presents the difference between the extracted center position and the input position in the simulation.
The maximum difference ($\ang{;3.2}$) was assigned to the systematic error.
Although there are variations of brightness on the Moon that are not due to the Moon phase (e.g., \cite{beaming}), their effects are much smaller than those due to the Moon phase. It is worthwhile to consider them in the future study.

Table~\ref{systematics} presents a summary of the systematic uncertainties.
We calculate the square root of the quadrature sum of all errors.
In total, we assigned a systematic error of $\ang{;3.2}$ to the pointing calibration using the Moon.

\section{Conclusions}


We calibrated the pointing of the GroundBIRD telescope based on Moon observation data. The angular response of the Moon was modeled with respect to the Moon phase as well as the beam of the telescope. We modeled the pointing shift due to the tilts of the elevation and azimuth axes as well as the collimation offsets and the encoder offsets. We obtained the optimized pointing model using them. The residuals of the calibrated pointing from the true pointing were less than $\ang{;1.6}$, whereas our requirement was $\ang{;4.7}$. The root mean squares were $\ang{;0.6}$ and $\ang{;0.5}$ for the elevation and azimuth, respectively.Consequently, we successfully achieved an uncertainty of $\ang{;3.3}$ including all systematic uncertainties, which is lower than our requirement. For the CMB telescope, this was the first attempt at using pointing calibration with the Moon for a beam width of $\ang{;36}$ against the Moon’s angular size of $\ang{;30}$. Thus, we realized sufficient pointing precision.

We can frequently observe the Moon at high elevations. The orbital period of the Moon (monthly) is much shorter than that of planets ($\sim$ 10 years). In addition, the observation of the Moon with the high signal-to-noise ratio allows us to perform the unbinned likelihood fit to extract the center position of the Moon. 
It introduces the advantage to reduce the effect of atmospheric fluctuation.
In addition, we avoid a degradation of an angular resolution of detector response due to the binning of the data for the astronomical point sources. We discussed the systematic uncertainties related to the Moon. The non-uniformity of the Moon’s brightness was assumed the most conservative case. Nevertheless, it was one order of magnitude lower than the beam width. In conclusion, the established method in this study is applicable to other CMB telescopes whose beam width is sub-degrees.

\section*{Acknowledgment}
This work is supported by JSPS KAKENHI under grant numbers JP21H04485, JP20KK0065 and JP21H04485, JSPS Core-to-Core Program number JPJSCCA20200003, and JSPS Bilateral Program numbers JPJSBP120219943 and JPJSBP120239919. 
YS also acknowledges JP21J20290. 
KL and EW also acknowledge 2022R1A2B5B02001535 by NRF. 

\vspace{0.2cm}


\let\doi\relax


\begin{thebibliography}{9}

\bibitem{Kami}
M. Kamionkowski, A. Kosowsky, and A. Stebbins
Phys. Rev. Lett. 78, 2058 (1997)
\doi{https://doi.org/10.1103/PhysRevLett.78.2058}

\bibitem{Zal}
M. Zaldarriaga and U. Seljak,
Phys. Rev. D 55, 1830 (1997)
\doi{https://doi.org/10.1103/PhysRevD.55.1830}

\bibitem{Staro}
A. A. Starobinsky,
JETP Lett. 30, 682-685 (1979)





\bibitem{Brout}
R. Brout, F. Englert, E. Gunzig,
Annals of Physics 115, 78 (1978)
\doi{https://doi.org/10.1016/0003-4916(78)90176-8}

\bibitem{Starobinsky}
A. Staroninsky
Phys. Lett. B 91, 990102 (1980)
\doi{https://doi.org/10.1016/0370-2693(80)90670-X}

\bibitem{Kazanas}
D. Kazanas, 
Astrophysical Journal, Part 2 - Letters to the Editor, vol. 241 (1980)
\doi{https://doi.org/10.1086/183361}

\bibitem{Sato}
K. Sato,
Monthly Notices of the Royal Astronomical Society, 195, 467 (1981)
\doi{https://doi.org/10.1093/mnras/195.3.467}

\bibitem{Guth}
A. Guth, S. Pi,
Phys. Rev. Lett, 49, 1110, (1982)
\doi{https://doi.org/10.1103/PhysRevLett.49.1110}

\bibitem{Linde}
A. Linde,
Phys. Lett. B, 108, 389, (1982)
\doi{https://doi.org/10.1016/0370-2693(82)91219-9}

\bibitem{Albrecht}
A. Albrecht, P. Steinhardt,
Phys. Rev. Lett. 48, 1220 (1982)
\doi{https://doi.org/10.1103/PhysRevLett.48.1220}






\bibitem{PB_point}
The POLARBEAR Collaboration et al 2019 arXiv e-prints
\doi{https://arxiv.org/abs/1910.02608}


\bibitem{QUIET_point}
C. Bischoff et al., ApJ 768 9 (2013)
\doi{https://doi.org/10.1088/0004-637X/768/1/9}


\bibitem{Planck_point}
P. Ade et al., [Planck Collaboration]
A\&A 571, A1 (2014)
\doi{https://doi.org/10.1051/0004-6361/201321529}

\bibitem{PB}
The POLARBEAR Collaboration et al 2017 ApJ 848 121
\doi{https://doi.org/10.3847/1538-4357/aa8e9f}

\bibitem{astropy2013}
R. Thomas et al.,
A\&A 558, A33 (2013)
\doi{https://doi.org/10.1051/0004-6361/201322068}

\bibitem{astropy2018}
A. Price-Whelan et al., [The Astropy Collaboration]
AJ 156 123 (2018)
\doi{https://doi.org/10.3847/1538-3881/aabc4f}


\bibitem{GB}
O. Tajima, J Choi, M. Hazumi, H. Ishitsuka, M Kawai, M. Yoshida,
Proc. SPIE 8452, Millimeter, Submillimeter, and Far-Infrared Detectors and Instrumentation for Astronomy VI, 84521M (2012)
\doi{https://doi.org/10.1117/12.925816}

\bibitem{CLASS}
Z. Xu et al., ApJ 891 134 (2020)
\doi{https://doi.org/10.3847/1538-4357/ab76c2}


\bibitem{Komine}
J. Komine, 
master thesis, Kyoto University (2019).

\bibitem{mirror1}
Y. Mizugutch, M. Akagawa, H. Yokoi,
IEEE Antennas and Propagation Society International Symposium (1976) \doi{https://doi.org/10.1109/aps.1976.1147539}

\bibitem{mirror2}
C. Dragone,
The Bell System Technical Journal ( Volume: 57, Issue: 7, September 1978)
\doi{https://doi.org/10.1002/j.1538-7305.1978.tb02171.x}


\bibitem{mirror3}
H. Tran, A. Lee, S. Hanany, M. Milligan, T. Renbarger,
Appl Opt. 2008 Jan 10;47(2):103-9.
\doi{https://doi.org/10.1364/ao.47.000103}

\bibitem{MKID}
P. K. Day, H. G. LeDuc, B. A. Mazin, A. Vayonakis, J. Zmuidzinas,
Nature 425, 817–821 (2003)
\doi{https://doi.org/10.1038/nature02037}

\bibitem{hybrid_MKID}
S. J. C. Yates, J. J. A. Baselmans, A. Endo, R. M. J. Janssen, L. Ferrari, P. Diener, A. M. Baryshev
Appl. Phys. Lett. 99, 073505 (2011)
\doi{https://doi.org/10.1063/1.3624846}

\bibitem{antenna}
D. F. Filipovic, S. S. Gearhart, and G. M. Rebeiz, IEEE Trans. Microwave
Theory Tech. 41, 1738 (1993)
\doi{https://doi.org/10.1109/22.247919}

\bibitem{GB_readout}
J. Suzuki, H. Ishitsuka, K. Lee, S. Oguri, O. Tajima, N. Tomita, E. Won 
Journal of Low Temperature Physics volume 193, pages562–569 (2018)
\doi{https://doi.org/10.1007/s10909-018-2033-x}

\bibitem{ishitsuka}
H. Ishitsuka, M. Ikeno, S. Oguri, O. Tajima, N. Tomita, T. Uchida,
Journal of Low Temperature Physics volume 184, pages424–430 (2016)
\doi{https://doi.org/10.1007/s10909-015-1467-7}

\bibitem{Jihoon}
J. Choi, 
GroundBIRD: A Telescope for the Cosmic Microwave Background Polarization Measurement, 
ph.D thesis, Korea University (2015).
\url{https://cmb.kek.jp/data/paper/jhchoi_thesis.pdf}

\bibitem{QUIET_Qband}
QUIET Collaboration et al 2011 ApJ 741 111
\doi{https://doi.org/10.1088/0004-637X/741/2/111}

\bibitem{ikemitsu}
T. Ikemitsu, 
master thesis, Kyoto University (2020).

\bibitem{T_model}
N. Nakai, M. Tsuboi, Y. Fukui,
Observation of The Universe $<$ 2 $>$ Radio Astronomy [Astronomy of The Series Modern] (2020)

\bibitem{moon_model}
C. A. Bischoff,
OBSERVING THE COSMIC MICROWAVE BACKGROUND POLARIZATION ANISOTROPY AT 40 GHZ WITH QUIET,
ph.D thesis, The University of Chicago (2010).

\bibitem{Kutsuma1}
H. Kutsuma, M. Hattori, R. Koyano, S. Mima, S. Oguri, C. Otani, T. Taino, O. Tajima,
Appl. Phys. Lett. 115, 032603 (2019)
\doi{https://doi.org/10.1063/1.5110692}

\bibitem{gao}
J. Gao, 
"The Physics of Superconducting Microwave Resonators",
ph.D thesis, California Institute of Technology (2008)

\bibitem{pointing_model}
J. Mangum, ALMA Memo Ser., 366 (2001)
(available online at http://legacy.
nrao.edu/alma/memos/html-memos/alma366/memo366.pdf)

\bibitem{QUIJOTE_phD}
D. Denis Tramonte
Using CMB, LSS and Galaxy Clusters as Cosmological Probes
ph.D thesis, Instituto de Astrofisica de Canarias (2010).

\bibitem{astropy_error}
PyERFA:documents
https://pyerfa.readthedocs.io/en/latest/api/erfa.moon98.html\#erfa.moon98

\bibitem{beaming}
J. S. V. Lagerros,
Astronomy and Astrophysics, v.332, p.1123-1132 (1998)

\end{thebibliography}

\appendix

\end{document}